# Color filter arrays based on dielectric metasurface elements

Jonas Berzins[a,b], Fabrizio Silvestri[b], Giampiero Gerini[b,c], Frank Setzpfandt[a], Thomas Pertsch[a], Stefan M. B. Bäumer[b]

[a]Institute of Applied Physics, Abbe Center of Photonics, Friedrich-Schiller-University Jena, Albert-Einstein-Str. 15, 07745 Jena, Germany; [b]Optics Department, TNO, Stieljesweg 1, 2628CK Delft, Netherlands; [c]Electromagnetic Group, Eindhoven University of Technology, Den Dolech 2, 5600MB Eindhoven, Netherlands

## ABSTRACT

Digital imaging has been steadily improving over the past decades and we are slowly moving towards a wide use of hyperspectral cameras. The color filter arrays being the key components of such systems, it is essential to develop a robust and scalable solution for an effective way of controlling the transmission of light. Nanostructured surfaces, also known as metasurfaces, offer a promising solution as their transmission spectra can be controlled by shaping the wavelength-dependent scattering properties of their constituting elements. Here we present, metasurfaces based on silicon nanodisks, which provide filter functions with amplitudes reaching 70-90% of transmission and well suitable for RGB and CMY color filter arrays, the initial stage before the further development of hyperspectral filters. We suggest and discuss possible ways of expansion of color gamut and improvement of color values of such optical filters.

**Keywords:** color filter, RGB, CMY, color filter arrays, silicon, nanodisk, metasurfaces

## 1. INTRODUCTION

A two-dimensional arrangement of nanoscale scatterers, a metasurface, has the capability to alter phase, polarization and amplitude of light. Such metasurfaces open new horizons in optics with the possibility to replace conventional optical components, such as: polarisers, microlenses and, most importantly, optical filters.

In recent years, there has been a great interest in the generation of structural colors by using dielectric nanostructures and metasurfaces. In contrary to plasmonic materials, dielectric structures provide a reasonable solution towards a lossless optical system, while maintaining the control of its spectral features [1-4]. In this context, there are many great achievements regarding the structural color generation in the reflection regime, so called color printing [5-7]. However, even-though this might be used in many applications, in the case of color filter arrays the structural colors are required to be visibly determined in transmission as it is with the color filters used in the conventional optical cameras. Taking this into account, some of the first demonstrations of color generation in transmission regime were done by using silicon (Si) nanowires [8-10], later structural colors were also shown by utilising shorter Si structures – nanodisks [11-13] and even subwavelength gratings, nanoholes [14]. As noticed, Si is the most frequently used high-index dielectric material in the visible spectral range, because of its' relatively low cost, compatibility with the complimentary metal-oxide-semiconductor (CMOS) process and minimum losses. However, in most of the presented cases the color quality (or the amplitude of filter function) is lower compared to the conventional dye-based color filters, suggesting that the filters could be further optimised. In this paper, we conduct a study on optimisation of color filters based on Si nanodisks, expanding our current knowledge on their potential in color filter arrays.

*jonas.berzins@uni-jena.de

## 1.1 Fundamentals of Spectral Tunability

At the resonance of nanostructures, described by Mie theory [15], we observe a strong extinction of the incoming electromagnetic field, which directly translates to a dip in the transmission spectrum. The resonant behaviour of plasmonic nanostructures, based on such materials as gold (Au) or silver (Ag), relies on free electron oscillations and due to its nature is accompanied by a significant optical loss. In comparison, the resonances of dielectric nanostructures are due to oscillations of bounded electrons, which do not induce Ohmic damping, providing minimal amount of losses and heating of the structures. By tuning of electric dipole (ED) and magnetic dipole (MD) resonances in dielectric nanostructures we have a high flexibility in control of light transmission [2]. The spectral position of MD resonance is considered to be approximately at the range of optical distance ($\lambda_{MD} = nD$)[3], so a focus is intended towards high refractive index dielectric materials. Even-though there is no perfect material, previously mentioned Si, as well as germanium (Ge) or gallium phosphide (GaP) are relatively good solutions [16]. In this case, the absorption and reflection is impinged at the resonant position and high transmission zones are observed outside of them [Fig. 1]. Such tunability allows to select particular ranges in the visible spectral range and create the colors required for the RGB and CMY color filter arrays.

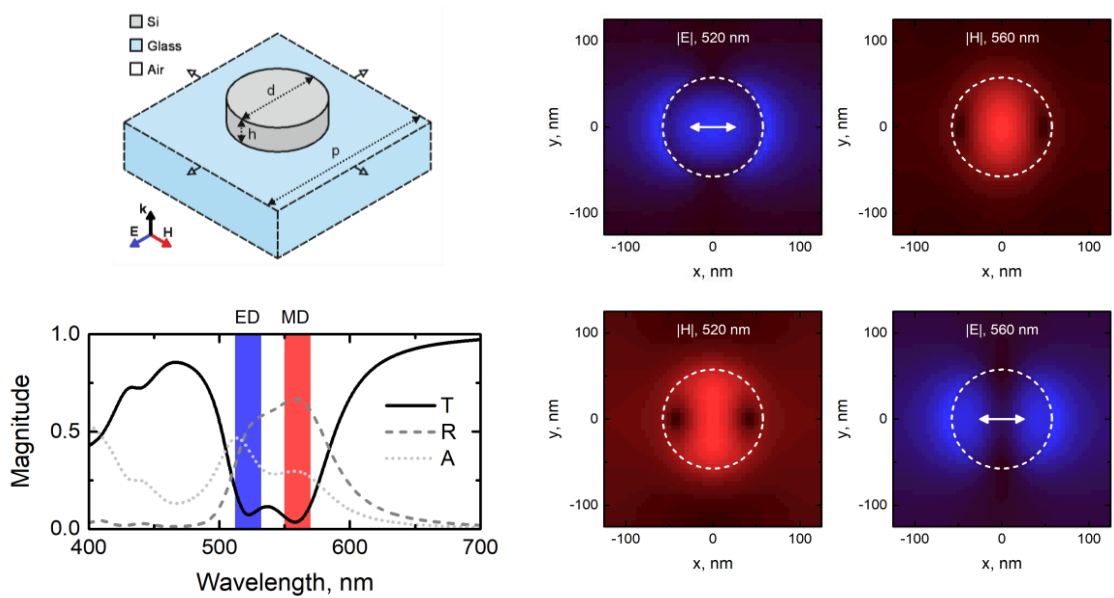

Fig. 1. Schematic representation of Si nanodisk array and its spectral properties: transmission (T), reflection (R) and absorption (A), when disks are 175 nm in height, 115 nm in diameter and distributed in a periodic array with a lattice constant (period) of 250 nm. The electric and magnetic field profiles of the disk at 520 nm and 560 nm wavelengths, shown on the right, correspond to electric dipole (ED) and magnetic dipole (MD) resonances.

## 1.2 Color Science

Imaging of basic optical cameras is based on the vision of a human eye. In the same way as it is with our color perception, the mechanism usually relies on three receptors, for example the well-known Bayer sensor based on red, green and blue (RGB) channels [17]. In order to identify the best filter function for a particular color, one has to define the color reference and the figure of merit.

Spectral composition of visible radiation can be characterised by the introduction of color matching functions, which is the base of the CIE Standard Colorimetric System [18, 19]. Each of the color matching functions ($S_x$, $S_y$, $S_z$) are linearly independent and are used to obtain tristimulus values ($X$, $Y$, $Z$) from the given spectral data (Fig. 2). To calculate these values, we multiply our simulation data by a normalised power distribution of day-light ($P_{D65}$), the illumination environment our camera is expected to be in, then integrate over the specified color matching functions, as shown in the first equation. This allows us to numerically identify a color from a measured spectra and then be able to accurately reproduce it.

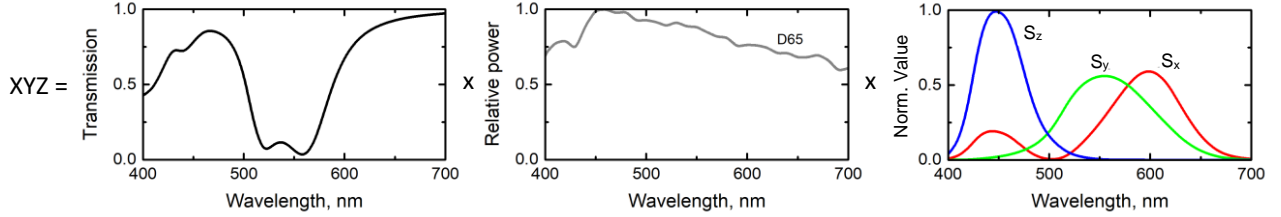

Fig. 2. Spectral data transformation into the coordinates of the CIE 1931 color space: multiplication of a generated spectra (left) by the relative power of chosen illuminant (center) and the integration of result over color matching functions (right).

$$X = \int T(\lambda) \times P_{D65}(\lambda) \times S_x(\lambda) d\lambda ,$$
$$Y = \int T(\lambda) \times P_{D65}(\lambda) \times S_y(\lambda) d\lambda , \quad (1)$$
$$Z = \int T(\lambda) \times P_{D65}(\lambda) \times S_z(\lambda) d\lambda .$$

Since the perceived color depends only upon the relative magnitudes, tristimulus values ($X$, $Y$, $Z$), we normalise them in order to define the chromaticity coordinates:

$$x = \frac{X}{X+Y+Z}, \qquad y = \frac{Y}{X+Y+Z}. \quad (2)$$

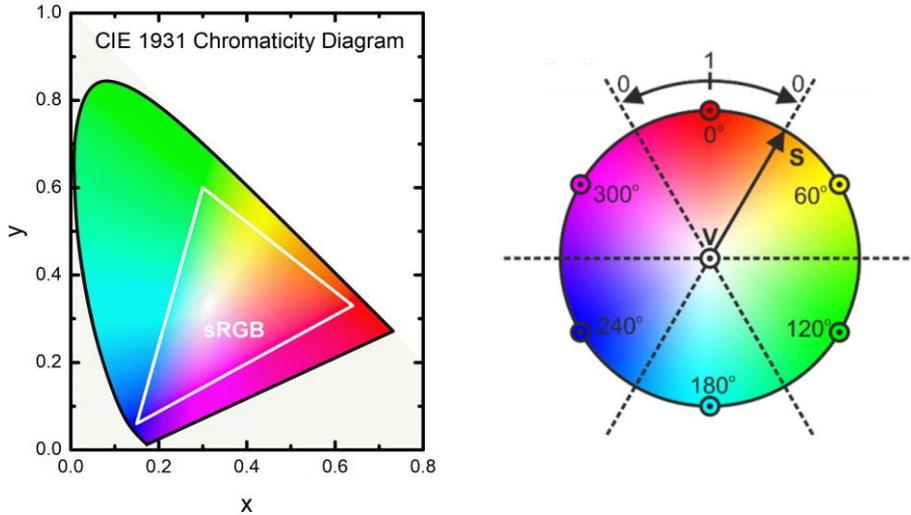

Fig. 3. The CIE 1931 chromaticity diagram with sRGB color space shown as a white triangle (left) and the transversal plane of the HSV color space with the schematic representation of the suggested $HSV_\xi$ figure of merit components (right).

The chromaticity diagram (Fig. 3) is commonly used to evaluate a color gamut. The irregular shape represents the limits of human vision, the outer line being equal to color bands of 1 nm width. The triangle inside the chromaticity diagram usually represents the RGB color space used by the device, in this instance sRGB color system is shown, because it is one of the most popular RGB working spaces. The assumption is that if the chromaticity of the color is within the boundary, then the color may be reproduced on that device, or may be represented by that color system. However, it should be noted that chromaticity diagram is only two-dimensional and does not include the value of luminance. Because of this, using only a chromaticity diagram to determine a color is incorrect solution, but it may help in the search of hyperspectral filters, relating the area of coverage to the prospect of the filters in hyperspectral imaging: the wider the gamut – the narrower the bands. The determination of the color requires a transformation to a particular color space. In our case we suggest the use the previously mentioned sRGB, but one can use different color spaces, as long as the colors are compared in-between for the search of the best, the values are valid. These values are usually achieved by multiplying the tristimulus values by the matrix of the RGB color space and the result can be later transformed into hue, saturation and value (HSV color space):

$$\begin{bmatrix} R \\ G \\ B \end{bmatrix}_{sRGB} = [M]_{sRGB} \begin{bmatrix} X \\ Y \\ Z \end{bmatrix} \rightarrow \begin{bmatrix} H \\ S \\ V \end{bmatrix} \qquad (3)$$

However, both color spaces use three values to identify a color (R, G, B or H, S, V), so it is relatively inconvenient to use those values as a figure of merit when comparing the color produced from the computed or measured spectra to the reference (ideal color) used in the case of color filter arrays. Taking this into account, here we suggest a simple method to assign the generated color to the closest color and then, depending on the distance from it, defined by the difference in hue, and the levels of saturation and value, provide a single merit:

$$HSV_\xi = \begin{cases} 0, & if\ \Delta H_\xi \geq \dfrac{\pi}{n} \\ \left(1 - \dfrac{n}{\pi} \times \Delta H_\xi\right) \times S \times V \end{cases}, \qquad (4)$$

where $n$ is the number of primary colors in the color scheme, $\Delta H_\xi = |H_\xi - H|$ is a difference of the hue from the hue of a specific color, $S$ is the saturation and $V$ is the value of a color or basically the amplitude of the signal. Color is indexed by ξ, so $H_R = 0$ rad, $H_G = 2\pi/3$ rad, $H_B = 4\pi/3$ rad, $H_C = \pi$ rad, $H_M = 5\pi/3$ rad, $H_Y = \pi/3$ rad, which results that only one of $HSV_\xi$ will have a positive value and this value can be considered as a figure of merit for a particular color and will be used in the following sections.

## 2. COLOR OPTIMISATION

The initial optimisation of dielectric metasurface elements for color filter arrays was done based on nanodisks made from crystalline silicon (c-Si), with wavelength dependent dispersion parameters [20], on top of the glass substrate ($n_{glass} = 1,46$), considering air ($n_{air} = 1$) as the covering medium (as shown in Fig. 1). The simulations were carried out by using commercial-grade simulator based on the finite-difference time-domain method [21]. The model was considered as an infinite system with periodic boundary conditions (PBC) on the sides and perfectly matched layers (PML) on the top and the bottom. In general case, monitors were placed to record the electric and magnetic fields as well as the transmission and reflection, which is done through integration of the collected power, as a function of wavelength, a normal-incidence plane wave source was used with the band-width covering the visible spectral range.

During the optimisation of filter functions, which correspond to the RGB and CMY colors, certain requirements coming from system engineering and nanostructure fabrication have to be met. Considering the primer, it is highly recommended to obtain a uniform height of structures throughout the color filter array, because variation in height requires multi-step e-beam lithography and etching process and more complicated implementation into an optical system. Considering the latter, the fabrication with e-beam lithography has limited resolution, in this case, due to a high aspect ratio, the minimum gap between the neighbouring disks and also the minimum size of the disk is set as 50 nm. Taking this into account a sweep through diameter (d) of nanodisks was made at a number of periods, ranging from 100 nm to 350 nm, and heights, ranging from 50 nm to 300 nm, in order to determine the best parameter window for RGB colors (Fig. 4).

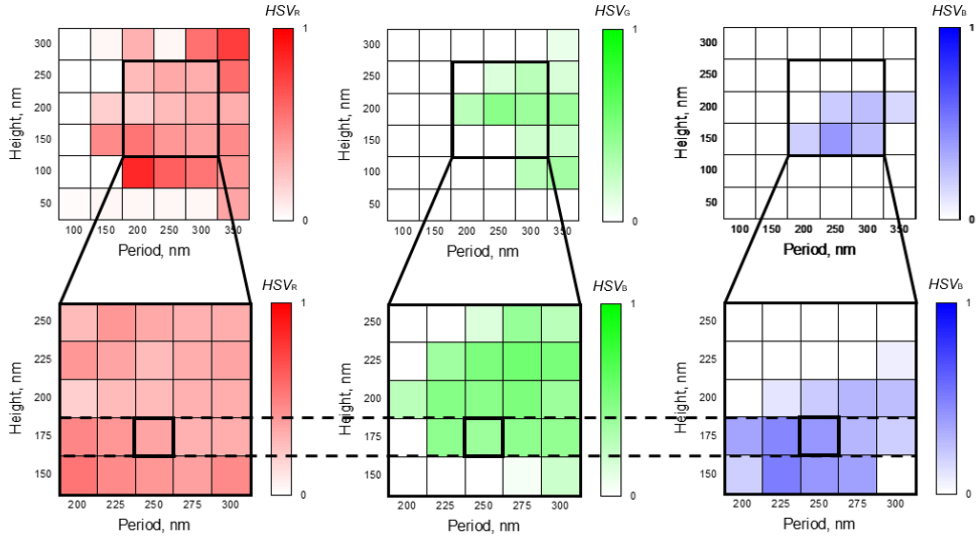

Fig. 4. Parameter window of c-Si nanodisks for optimal RGB colors. In each set of period and height a diameter sweep is done in order to acquire the best $HSV_\xi$ values. The optimal height of the stuctures is considered to be equal to 175 nm.

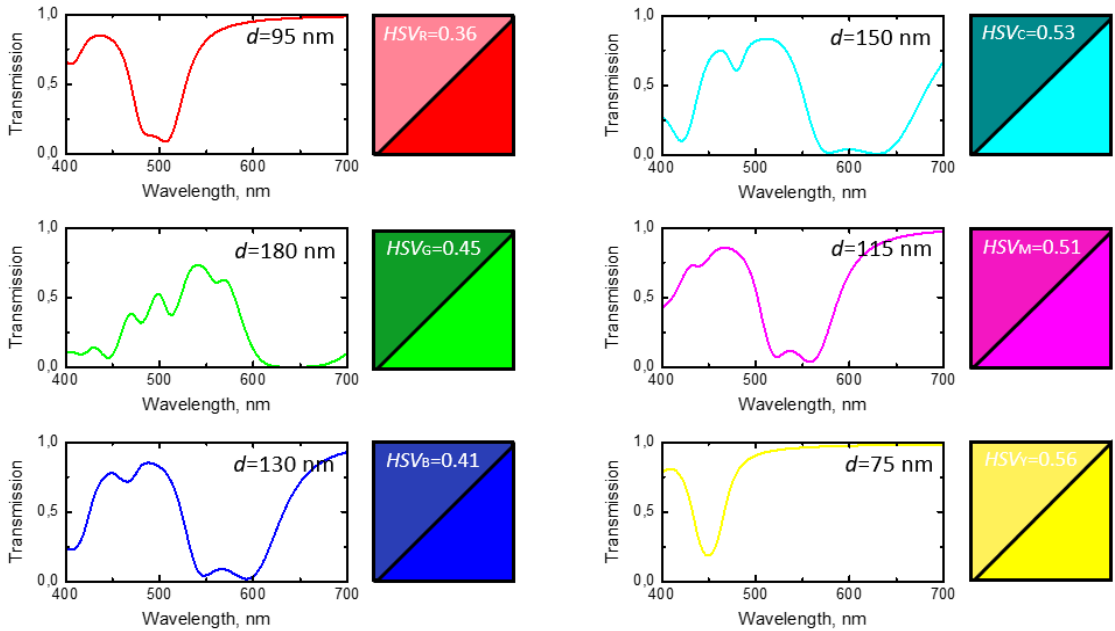

Fig. 5. RGB and CMY filter functions and their representation in colors achieved by tuning the physical properties of c-Si nanodisks on top of glass substrate, the spectra tranformed into color (upper triangle) is compared to the ideal color (lower triangle).

The structures with optimised height (175 nm) and period of 250 nm were further tuned in order to achieve the best filter functions: in the case of RGB, the diameter of nanodisk for red color is equal to 95 nm, green – 180 nm, blue – 130 nm. In the case of CMY, caen – 150 nm, magenta – 115 nm, yellow – 75 nm (Fig. 5). It has to be noted that the filter function optimisation is carried out considering only the visible spectral range, being only presumptuous of the behaviour in the ultraviolet and infrared spectral ranges. Despite this, such assumption should not limit the filters application-wise due to the fact that the usually exploited cameras consists of additional IR-cut filter on the lens, while the effect of the ultraviolet illumination due to higher order modes should be looked at as negligible.

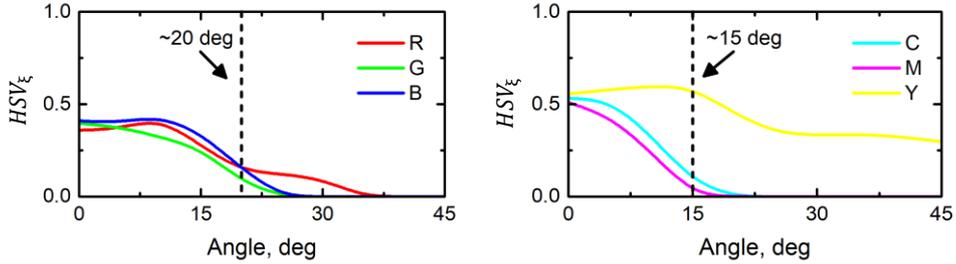

Fig. 6. Angle tolerance of RGB and CMY color filters based on uniform c-Si nanodisk arrays. The height of the disks is considered optimal at 175 nm and the period of the disks are chosen to be equal to 250 nm.

In addition to previously mentioned requirements, angle tolerance is an important system specification. In the real-life applications the incident beam arrives as a cone of rays rather than a single ray, this cone might be incident at the normal incidence or at particular angle. Either-way, the filter must possess a low angle dependence or angle independence feature. For this simulation we selected previously optimised nanodisks and observed the transmission behaviour by changing the angle of incidence. As a result, we conclude that the metasurfaces resist well to the change of an angle, while the RGB colors can be discriminated till 20 degrees and CMY – 15 degrees angle of incidence.

## 3. COMPARISON OF DIFFERENT TYPES OF SILICON

As mentioned in the introductory section, silicon is the most frequently used high-index dielectric material. Despite considering silicon almost lossless in the infrared spectral range, in the visible range it is not (Fig. 7, left-bottom). According to the Kramers-Kronig relations [22] any dispersion of permittivity is related to dissipation due to exciton and phonon resonances. This fundamental trade-off between high refractive and absorption eventually sets the limit to the performance of all-dielectric metasurface. As the dispersion parameters of Si are highly dependent on the crystallinity of the material, in addition to previously presented results by using c-Si, in the right of Fig. 7, we take a look at other types of silicon: amorphous silicon (a-Si) and hydrogenated amorphous silicon (a-Si:H).

In comparison to others, the growth of c-Si on top of foreign substrate is the most complicated process. So it is worth looking into alternatives: the use of a-Si simplifies the fabrication and recently there has also been a study on the use of hydrogenated amorphous silicon (a-Si:H) for metasurface-based CMY color filters [12] as the losses diminish compared to the initial of a-Si, but are still impactful in the visible spectral range. The dispersion properties of a-Si and a-Si:H were measured, while the c-Si were taken from the database [20]. Within the scope, high-temperature annealing of deposited a-Si structures can be considered in order to reach poly-crystalline state of matter, however the achievable dispersion properties may vary depending on the variety properties of the fabrication process such as temperature and duration, also quality of a-Si. In other words, despite the expectations being set, this may result into crystallinity state of the material with losses between a-Si and c-Si, but for the deterministic analysis a proper experimental data is required, thus is not considered at this point.

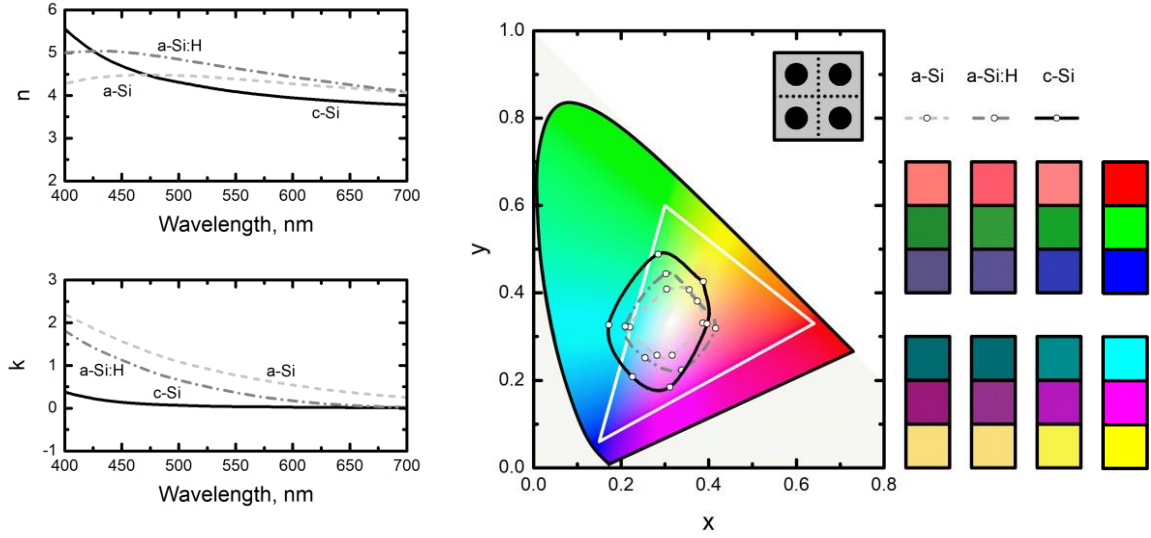

Fig. 7. The dispersion parameters (left) and effectiveness of color generation (right) of different type of silicon nanodisk: black solid line – c-Si, dark grey dot-dashed line – a-Si:H and dashed light grey line – a-Si. The optimised filter functions represented by colors visually compared to the ideal RGB and CMY colors.

The results of simulation of varying 175 nm height nanodisks' diameter from 50 nm to 200 nm in a periodic array with 250 nm period suggests that RGB and CMY colors can be successfully discriminated independently of the silicon type. Of course, due to smaller losses the value and saturation level of colors are higher in c-Si, except red, because in case of a-Si and a-Si:H the blue region is diminished to losses while playing a role of increasing the value of red color. In addition to that, the gamut achieved by using a-Si is 33% of the one achieved with c-Si, while the use of a-Si:H increases the gamut almost twice, to 61% of the c-Si. However, even with the use of c-Si the covering area reaches only 45% of the sRGB gamut. According to this, the following sections will be covering the further expansion of the gamut by considering c-Si as the basis material of the investigated structures.

## 4. EXPANSION OF COLOR GAMUT

Single layer of nanostructures provides a single filter function, but it is well known that putting multiple filters next to each other can transform white light source into variety of bands, the transmission of the filter system being equal to the multiplication of the filter functions. The stacking of metasurfaces [23] can be used to expand the gamut. Also the mixing of the structures [24] is a promising way to add an additional degree of control and by doing this improve the filter functions.

### 4.1 Vertical stacking of metasurfaces

In theory the number of filter layers can go to infinity, but in practical application any additional layer is equal to additional steps in the fabrication, thus in the simulations the number of filters (layers) is changed in a limited range, from 1 to 5. The filter functions are taken from the previously generated simulation data to fit the requested color function, in other words for two-layer system to multiply any two filter functions to get the highest possible HSV values for RGB and CMY colors and repeat this for a 3, 4 and 5 layers (Fig. 8). In our simulations the coupling between the layers is not considered and the interference effect is considered as negligible.

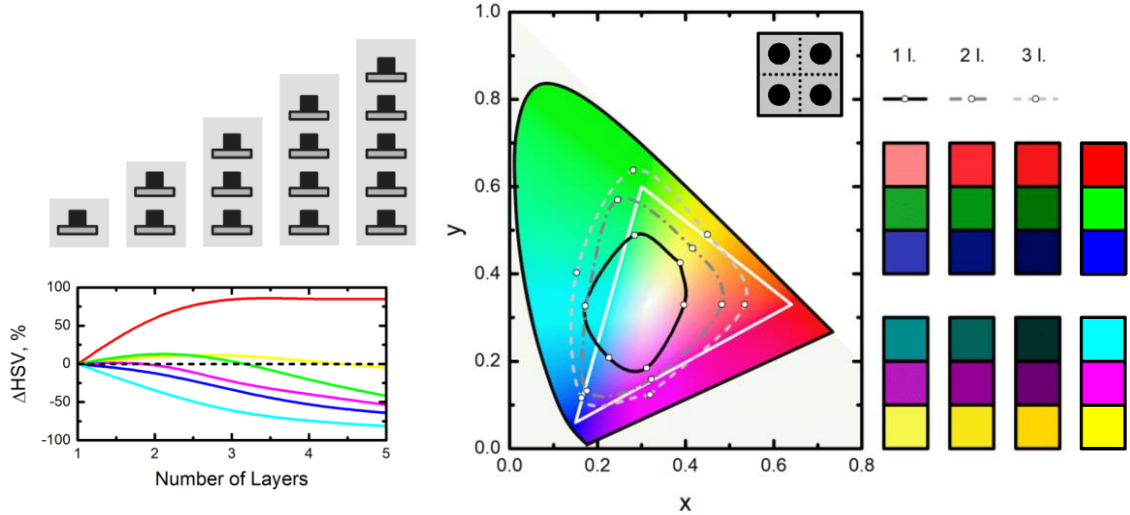

Fig. 8. The effect on colors by vertical stacking of c-Si metasurfaces: the change of HSV value (left), increase of gamut and the change in color can be observed (right). Colors are compared to the ideal RGB and CMY colors.

This multi-layer approach gives a rise in gamut: by using two layers the gamut is increased by 238%, using 3 layers – even 352% of the initial single layer gamut (106% and 157% compared to sRGB, respectively). Even-though the color gamut eventually exceeds the sRGB gamut, this is achieved at the cost of the luminance for most of the cases, except red, where quite clear improvement of color is seen (the HSV value is doubled going from one to three-layer system).

### 4.2 Horizontal mixing of nanostructures

As a solution for spectral combination can be considered variation of nanodisks in a single layer – single metasurface [24]. Due to small period (250 nm) compared to the wavelengths, high coupling between the nanodisks is observed, but this is controllable if the higher order modes are not present within the investigated spectral range.

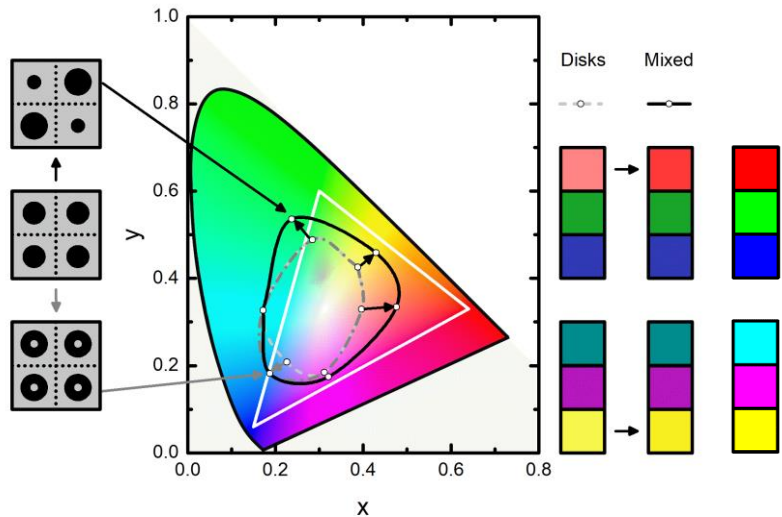

Fig. 9. The effect on colors by horizontal mixing of c-Si metasurfaces. Nanodisks of different sizes and rings are introduced. Optimal colors are compared the ideal RGB and CMY colors.

Despite the gamut expansion, only red and yellow are significantly improved (Fig. 9). The change in other colors is minimal. The introduction of rings improve the gamut but does not contribute to the increase the color quality. Finally, the optimal setup for RGB color filter array is suggested: red color – mixing of 80 nm and 105 nm diameter nanodisks, green –180 nm nanodisks, blue – 130 nm nanodisks. The setup for CMY color filters array is the following: caen – 150 nm nanodisks, magenta – 115 nm nanodisks, yellow – mixing of 70 nm and 80 nm nanodisks. The period of the disks is 250 nm and the height is 175 nm. Filter functions are shown in Fig. 10.

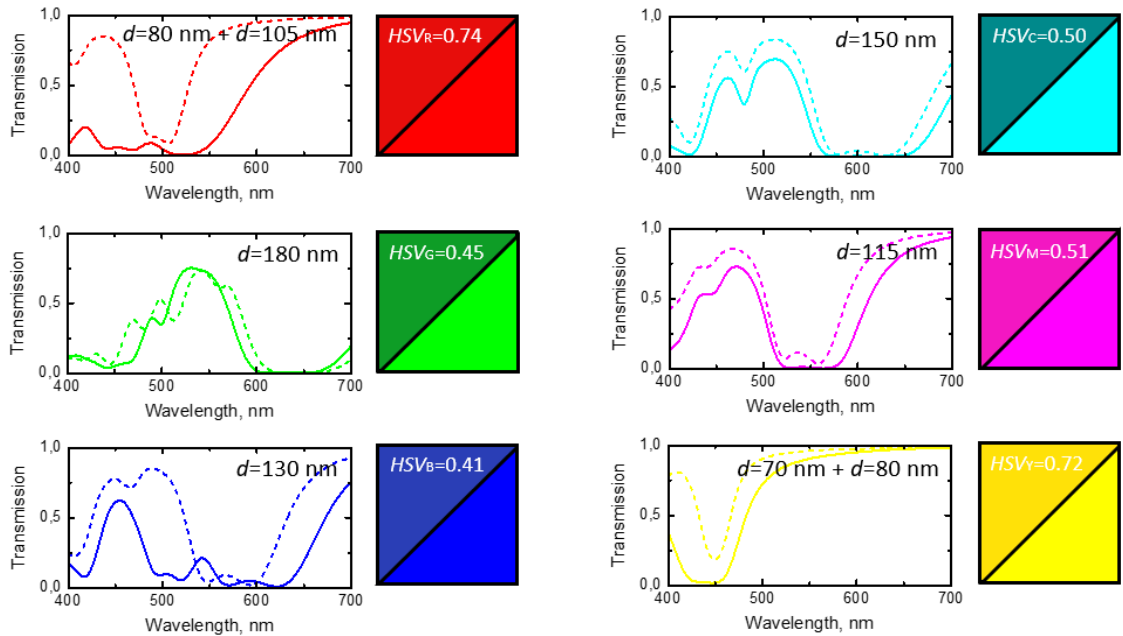

Fig. 10. Optimised color filter functions: solid line - using simple array of nanodisks, dashed line – array of mixed nanodisks. The given numbers at the top graph suggest the size of constituents for optimal color value. the spectra tranformed into color (upper triangle) is compared to the ideal color (lower triangle).

## 5.   OUTLOOK AND DISCUSSION

Step by step optimization of metasurface based components of color filter array was carried out. For this matter a new figure of merit for spectra comparison to the ideal colors was introduced. Such approach lead into optimization of the c-Si nanodisk metasurfaces, their filter function amplitudes reaching 70-90% of transmission, while the color filter maintains the color discrimination property of the selected color up to 20 degrees for RGB and 15 degrees for CMY color filter array configurations. Different types of silicon (c-Si, a-Si, a-Si:H) can be successfully used for the discrimination of RGB and CMY colors, however, as predicted, for amorphous cases the quality of colors is lower due to higher absorption. Eitherway, filtering properties can be further enhanced by the vertical stacking, extending the gamut over the limits of sRGB, or horizontal mixing, highly improving the color quality without impinging additional technological steps.

Even-though this study was performed as the initial step before the construction of more complicated hyperspectral filters, the results also suggest advantages in comparison to conventionally used dye-based RGB and CMY color filters, which degrade under ultraviolet illumination and high temperature environments, as the scalability of nanostructured filters is extremely promising: a single unit of an array already possess color property, which means a pixel could be scaled down below 1 μm and the sensor could be flexibly customised towards hyperspectral imaging.

## ACKNOWLEDGMENTS

This project has received funding from the European Union's Horizon 2020 research and innovation programme under the Marie Sklodowska-Curie grant agreement No 675745.